\documentclass[10pt,twocolumn,letterpaper]{article}
\usepackage[inline]{enumitem}
\usepackage{cvpr}
\usepackage{graphicx}
\usepackage{amsmath}
\usepackage{amssymb}
\usepackage{booktabs}
\usepackage[ruled,vlined]{algorithm2e}
\usepackage{epsfig}
\usepackage{physics}
\usepackage{comment}
\usepackage{subcaption}
\usepackage{framed}
\usepackage{multirow}
\usepackage{xcolor,soul}
\usepackage{latexsym,url}
\usepackage{stackengine} 

\usepackage[pagebackref,breaklinks,colorlinks]{hyperref}

\usepackage[capitalize]{cleveref}
\crefname{section}{Sec.}{Secs.}
\Crefname{section}{Section}{Sections}
\Crefname{table}{Table}{Tables}
\crefname{table}{Tab.}{Tabs.}

\def\cvprPaperID{14411} 
\def\confName{CVPR}
\def\confYear{2024}

\definecolor{todocolor}{RGB}{200,120,120}
\sethlcolor{todocolor}

\def\etal{{et.~al}}

\def\gbc{\textsc{gbc}\xspace}
\def\gb{\textsc{gb}\xspace}

\def\myarch{{FocusMAE}\xspace}

\newcommand{\myfirstpara}[1]{\noindent \textbf{#1.}}
\newcommand{\mypara}[1]{\vspace{0.1em} \myfirstpara{#1}}

\newcommand{\beginsupplement}{%
        \setcounter{table}{0}
        \renewcommand{\thetable}{S\arabic{table}}%
        \setcounter{figure}{0}
        \renewcommand{\thefigure}{S\arabic{figure}}%
     }

\begin{document}

\title{FocusMAE: Gallbladder Cancer Detection from Ultrasound Videos with \\ Focused Masked Autoencoders}

\author{Soumen Basu\textsuperscript{1}%
\thanks{~Soumen is currently affiliated to Samsung R\&D Institute Bangalore}~~\thanks{~Joint first authors}~, 
Mayuna Gupta\textsuperscript{1~\textdagger}, Chetan Madan\textsuperscript{1}, Pankaj Gupta\textsuperscript{2}, Chetan Arora\textsuperscript{1} \\
\textsuperscript{1} IIT Delhi \qquad 
\textsuperscript{2} PGIMER, Chandigarh\\
}

\maketitle

\begin{abstract}
    In recent years, automated Gallbladder Cancer (GBC) detection has gained the attention of researchers. Current state-of-the-art (SOTA) methodologies relying on ultrasound sonography (US) images exhibit limited generalization, emphasizing the need for transformative approaches. We observe that individual US frames may lack sufficient information to capture disease manifestation. This study advocates for a paradigm shift towards video-based GBC detection, leveraging the inherent advantages of spatiotemporal representations. Employing the Masked Autoencoder (MAE) for representation learning, we address shortcomings in conventional image-based methods. 
    We propose a novel design called \myarch to systematically bias the selection of masking tokens from high-information regions, fostering a more refined representation of malignancy. Additionally, we contribute the most extensive US video dataset for GBC detection. We also note that, this is the first study on US video-based GBC detection. We validate the proposed methods on the curated dataset, and report a new SOTA accuracy of 96.4\% for the GBC detection problem, against an accuracy of 84\% by current Image-based SOTA -- GBCNet and RadFormer, and 94.7\% by Video-based SOTA -- AdaMAE. We further demonstrate the generality of the proposed \myarch on a public CT-based Covid detection dataset, reporting an improvement in accuracy by 3.3\% over current baselines. Project page with source code, trained models, and data is available at: \href{https://gbc-iitd.github.io/focusmae.html}{\texttt{https://gbc-iitd.github.io/focusmae}}. 
\end{abstract}

\begin{figure}
    \centering
    \begin{subfigure}[b]{\linewidth}
        \centering
        \includegraphics[width=0.95\linewidth]{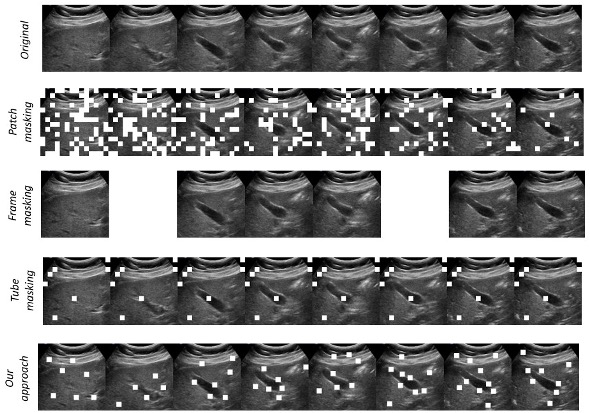}
        \caption{}
        \label{fig:teaser_a}
    \end{subfigure}
    
	\begin{subfigure}[b]{\linewidth}
		\centering
		\includegraphics[width=0.95\linewidth]{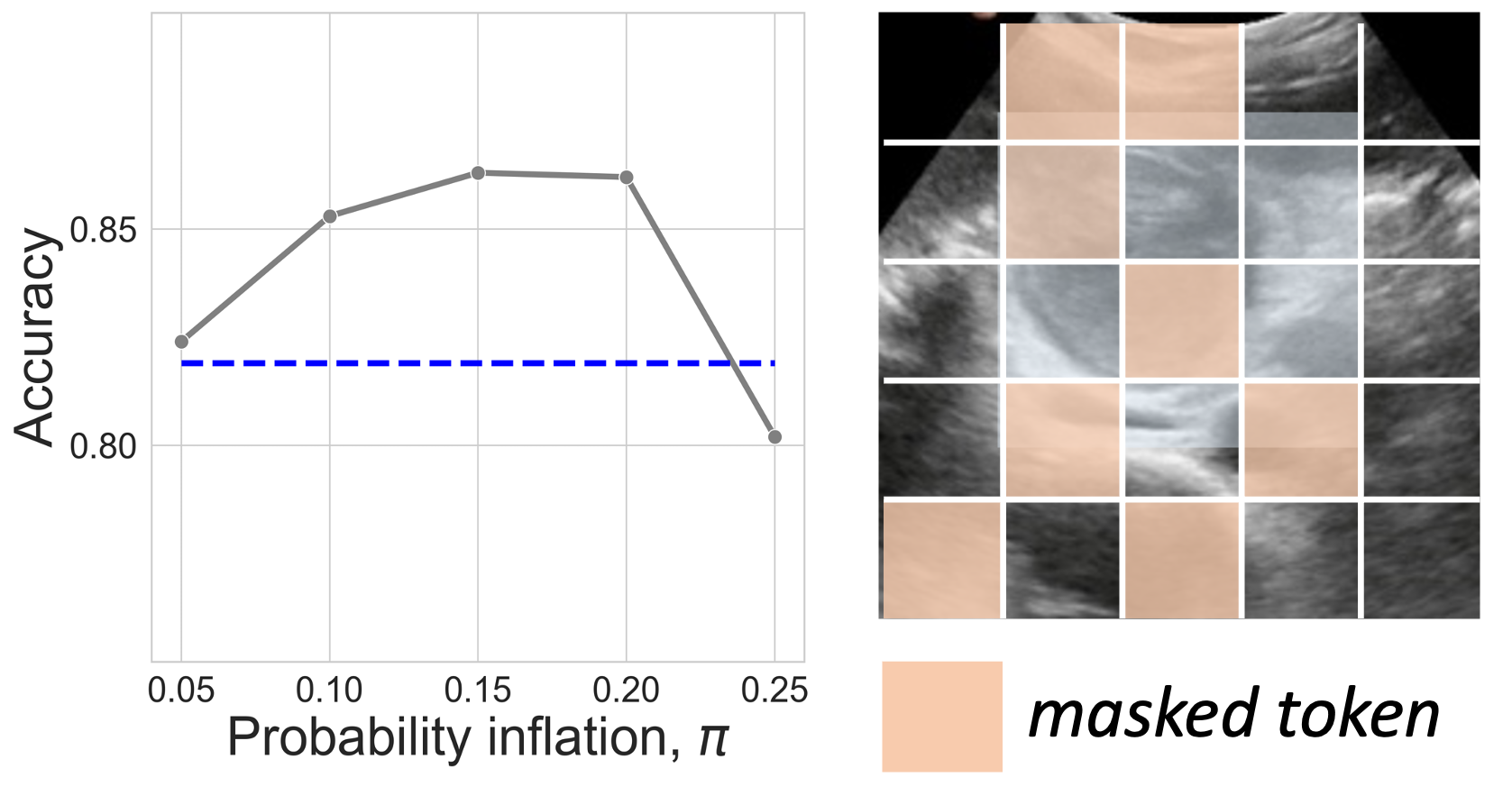}
		\caption{}
		\label{fig:teaser_b}
	\end{subfigure}
    \caption{(a) Masking strategy of FocusMAE in comparison to existing random patch \cite{maest}, frame \cite{wei2022masked}, tube \cite{videomae} masking. Our approach selects more tokens from the semantically meaningful regions with a small number of background tokens for masking. (b) Inflating the masking probability of the tokens which spatially lie within the object region (gray region) by $\pi$ increases the accuracy. However, excessive masking of the object region degrades performance. Blue line: accuracy of the original random masking. }
    \label{fig:teaser}
\end{figure}

\section{Introduction}

\myfirstpara{Gallbladder Cancer (GBC)}
Lately, automated GBC detection has drawn an increased interest from the researchers \cite{basu2022surpassing, basu2023radformer, chang2022ct, kinoshita2023deep}. GBC is difficult to detect at an early stage \cite{howlader2017seer}, and surgical resection becomes infeasible for most patients as the disease gets detected at a late stage. As a result, the disease shows bleak survival statistics. The 5-year survival rate for patients with advanced GBC is only 5\%, and the mean survival time is six months \cite{randi2006gallbladder, gupta2021locally}. Hence, early detection of GBC is crucial for timely intervention and improving the survival rate \cite{hong2014surgical}. 

\mypara{Ultrasound (US) for GBC Detection}
US has been the preferred non-invasive diagnostic imaging modality owing to its low cost, accessibility, and non-ionization. Often, it is the sole imaging performed on patients with abdominal diseases in low-resource countries. However, unlike benign afflictions like stone or polyp, identifying signs of malignancy from routine US is challenging for radiologists \cite{gupta2020imaging,gb-rads-paper}. GBC may advance silently if it remains undetected. Thus, it is imperative to identify GBC from US at an early stage. 

\mypara{Automated Detection of GBC}
Detecting GBC from US images using Deep Neural Networks (DNNs) is challenging. US images often have low quality due to sensor issues, causing biases in DNNs and making it hard to pinpoint the gallbladder (GB) region accurately \cite{basu2022surpassing}. The handheld nature of the probe also means the views are not aligned, adding to the challenge. Malignant cases, unlike non-malignant ones with clear anatomy, are difficult to detect due to the lack of a distinct GB boundary or shape and the presence of masses. While there are recent efforts to circumvent the challenges of US for accurate GBC detection \cite{basu2022surpassing, basu2023radformer, basu2022unsupervised}, these techniques are primarily image-based. Due to the challenges discussed earlier, single images may lack unambiguous features for malignancy detection. We also observe in our experiments that the image-centric methods do not generalize well to unseen datasets. In response, we argue in favor of a paradigm shift to video-based GBC detection from US. Notably, video-based GBC detection from US has not been attempted in the literature. 

\mypara{Masked Autoencoders (MAEs)}
Recently, MAEs \cite{maest, videomae, adamae, mgmae,videomaev2} have emerged as a promising representation learning technique for vision-related tasks. The idea behind MAE is to mask certain parts (also called tokens) of the input and then try to reconstruct the masked parts from the visible parts as a pretraining task. Usually, a Vision Transformer (ViT) \cite{vit} generates the embedding of the tokens. Mask sampling strategy plays a significant role in effectively learning using MAEs \cite{videomae,maest}. Currently, a random masking strategy is adopted in most MAE approaches \cite{videomae, videomaev2, maest}. For random masking in videos, patch masking \cite{maest}, frame masking \cite{wei2022masked, qian2021spatiotemporal}, or tube-based masking (dropping tokens at the same spatial location across a few consecutive frames) \cite{videomae} are popularly used. Tube-based masking strategy is considered to be better at preventing information leakage arising from redundancy in the time dimension. However, studies suggest that a single masking strategy may not fit all datasets due to the diversity of scenes, acquisition conditions, and high/ low spatiotemporal information regions in videos \cite{adamae}. For example, Video-MAE \cite{videomae} achieves the best action classification on the SSv2 dataset  \cite{ssv2} with random tube masking. For Kinetics-400, MAE-ST attains the best performance through random patch masking. \cref{fig:teaser_a} shows examples of different masking strategies. 

\mypara{Our Proposal}
In US videos, GB and malignant regions typically occupy a tiny portion. Notice that these are high information regions as opposed to non-GB portions of the frames, which are low information regions. Thus, random masking (uniform distribution) is not conducive to learning effective representations of malignancy. Few recent approaches suggest using an adaptive mask sampling strategy for more meaningful semantic representation \cite{adamae, mgmae}. MGMAE \cite{mgmae} suggests using object motions to guide the mask sampling. AdaMAE \cite{adamae} exploits a policy gradient optimization strategy by maximizing the expected token reconstruction error in order to boost the sampling probability of the tokens belonging to the objects. Since the organs are mostly stationary in US videos or CT volumes, the motion-guided strategy is not applicable to our case. On the other hand, our experiments show that AdaMAE does not perform significantly better than VideoMAE. By focusing solely on reconstruction error, the model may underrepresent crucial features or patterns within the data. In contrast, we adopt a simple strategy, \myarch, in sampling effective masks in the MAE pipeline. We identify candidate high-information regions, and bias the sampling strategy with these region-priors to sample the masking tokens from these focused candidate regions. By using a stronger masking on the high information regions, and reconstructing these tokens, \myarch learns a more refined representation. 

\mypara{Contributions} 
The key contributions of this work are:
\vspace{-0.5em}
\begin{enumerate}[label=\textbf{(\arabic*)}]
\itemsep-0.6em
	\item We posit that existing SOTA techniques for GBC detection in US images exhibit suboptimal accuracy and generalization performance. Consequently, we advocate for a paradigm shift toward video-based GBC detection for US. Also, the problem of US video-centric detection of GBC with machine learning was not previously attempted in literature. We provide the first solution to the problem and present a strong baseline. 
	%
	\item Even though video-based GBC classification shows improvement over image-based methods in terms of accuracy, specificity, and sensitivity, we observe that the random masking in MAE presents opportunities for further improvement. Notably, the spatiotemporal regions indicative of malignancy typically constitute a small portion of the video. The random selection of masked tokens introduces redundant background information, necessitating a more systematic approach. To address the issue, we propose a novel design, \myarch, to systematically bias the masking token selection from the semantically meaningful candidate regions. As a result, the network is compelled to learn a more refined representation of GB malignancy while reconstructing the masked tokens. We report an accuracy of 96.4\% using our approach as against 84\% by the current SOTA of GBCNet \cite{basu2022surpassing} and Radformer \cite{basu2023radformer}\footnote{Both GBCNet and RadFormer gave an identical accuracy in our experiments. We confirmed that individual predictions were not identical.}.
	\item Our idea of focused masking is generic, and we validate the generality of the method by applying it to a public CT-based Covid identification task \cite{covidctmd}. We report an accuracy gain of 2.2\% by our method over the SOTA \cite{touvron2021training}.
	\item Concurrently, we curate the most extensive US video dataset available for GBC detection. We establish the dataset by adding 27 US video samples exhibiting GBC to the publicly available GBUSV dataset. The dataset will be made available to the community. 
    %
\end{enumerate}

\section{Related Work}
\myfirstpara{Deep Learning for GB related Diseases}
Several studies have leveraged DNNs to detect various GB conditions, including calculi, cholecystitis, and polyps, using diagnostic images. For instance, \cite{gbYolo} applied YOLOv3 to identify the GB and stones in CT images. \cite{gbPolyp} focused on GB segmentation and employed an AdaBoost classifier for polyp diagnosis. Meanwhile, \cite{gbPolyp2} concentrated on classifying neoplastic polyps in cropped gallbladder ultrasound (USG) images, utilizing an InceptionV3 model. \cite{jang2021diagnostic} employed ResNet50 to diagnose polypoidal lesions through endoscopic US.

\mypara{DNNs for GBC Detection}
Despite numerous studies on DNNs for gallbladder-related diseases, only a few have explored AI-based detection of GBC \cite{gupta2024applications}. Chang \etal \cite{chang2022ct} employed a UNet-based denoising to enhance the image quality of Low-Dose CT scans for characterizing GBC. 
In contrast, Basu \etal \cite{basu2022surpassing} introduced a CNN architecture called MS-SoP and a Gaussian blurring-based curriculum for efficient GBC detection in US images. Gupta \etal \cite{gbc-lancet} further studied the performance of MS-SoP in classifying different sub-types of GBC on a large prospective patient cohort. Basu \etal \cite{basu2022unsupervised} later utilized unsupervised contrastive learning to learn malignancy representations. On the other hand, \cite{basu2023radformer} exploits a transformer-based dual-branch architecture for accurate and explainable GBC detection. \cite{xgc} investigates application of transformers for differentiation of GBC with xanthogranulomatous cholecystitis. \cite{basu2023gall} further proposes DETR-based weakly supervised GBC detection. 
Gupta \etal \cite{gupta2023reliable} proposes a calibration metric and loss to calibrate the GBC detection models on small dataset.
Despite the above studies, we observe a notable gap in the literature regarding models for video-based GBC detection from US videos. This gap in research motivates the current work.

%

\mypara{Video-based Classification and Recognition}
Transformers have seen an influx over CNNs due to their superior performance. Transformers with combined spatiotemporal attention \cite{vivit}, hierarchical spatiotemporal attention \cite{videoswin}, and separable spatial and temporal attention \cite{vidtr, timesformer} are popular for video-based recognition or classification. 

\mypara{Masked Autoencoder for Videos}
MAEs have gained popularity for self-supervised video representation learning (SSL). \cite{maest, videomae} extend the MAE from image to video domain. \cite{omnimae} used a combined image and video-based MAE pipeline. On the other hand, \cite{qing2023mar} introduced running cell masking to reduce cost. Another study \cite{videomaev2} recommended masking decoder tokens as well. \cite{adamae} recommends an adaptive masking strategy instead of random masking. Some studies look for priors like motion trajectory \cite{mgmae,patrick2021keeping, Sun_2023_CVPR}. \cite{li2022semmae} recommends using semantic parts guided MAE. \cite{maskvit} introduces the usage of both spatial and spatiotemporal attention along with variable token masking ratio.

\begin{figure*}[t]
    \centering
    \includegraphics[width=\textwidth]{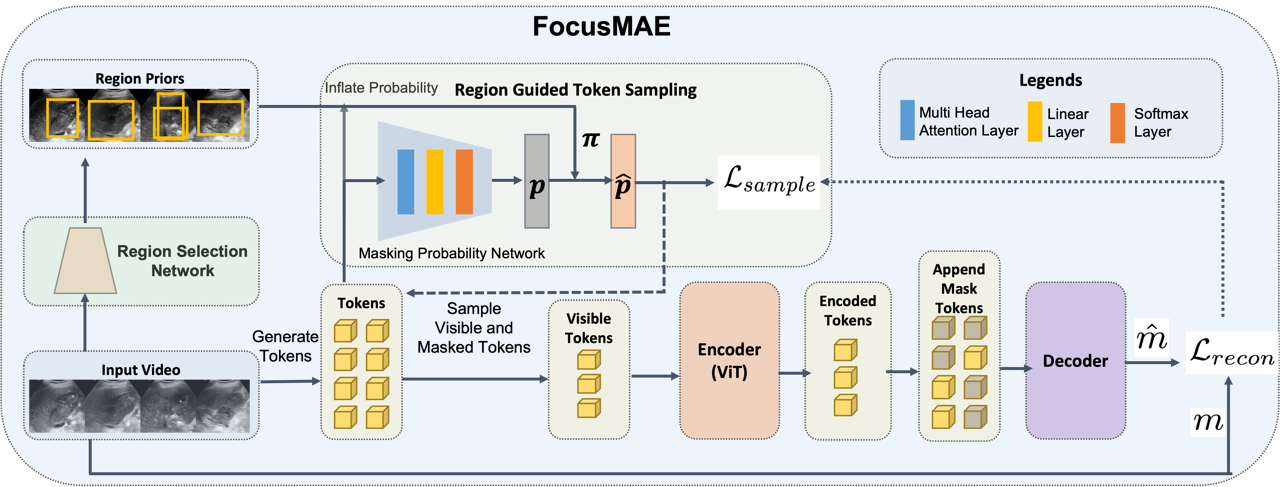}
    \caption{Overview of the proposed FocusMAE pipeline. Our design proposes guiding the masking tokens with the localization of the candidate focus regions containing high-information. The systematic biasing with focused high-information region priors helps to build a more meaningful reconstruction task for disease representation learning. }
    \label{fig:enter-label}
\end{figure*}

\section{Proposed Method}
\subsection{Object Priors in MAE}
Visual data often demonstrate sparser semantically meaningful information distribution dominated by the foreground objects. Current MAE techniques predominantly use random masking, which may result in sub-optimal results as the information may not be uniformly distributed. For the US videos, GBC often occupies a very humble portion of the frames. Random masking mostly biases the networks to learn representations of redundant backgrounds containing other organs or abdominal cavities. To alleviate the issue, we advocate exploiting the object location priors with high information density to enhance the representation learning in MAE. We show in \cref{fig:teaser_b} the preliminary evidence of potential advantages of boosting the masking token probability with object localization priors. We selected a random validation split containing about 20\% of our GB US Video dataset. We used the malignant ROI boxes provided in the dataset to specify object locations. We manually increased the masking probability of patches within the bounding box region for the data samples, and used them for self-supervised pretraining. We varied the probability boosting values, denoted by $\pi$, representing the increased probability for patches within the bounding box compared to those in the background. Our experiment reveals that an increase in the masking probability for patches within the bounding box, as opposed to random masking, leads to a noticeable enhancement in results. However, highly inflating masking probability for patches within the bounding box may compromise the integrity of the pretext task and result in performance degradation. These findings underscore the importance of recognizing that distinct image patches contribute differently to the learning of visual representations. Furthermore, the emphasis on reconstructing foreground objects with a balanced approach is crucial for optimal performance.

\subsection{\myarch Architecture}
\label{sec:method_subsample}
\mypara{Video Sub-sampling}
Video data contains temporal redundancy as the consecutive frames see a very high overlap in content. We sub-sample the videos to reduce the temporal redundancy. Assuming a video containing $F$ frames, we first sub-sample $\frac{F}{4}$ frames with a stride of $4$. Although the viewpoint in US frames can change very quickly, in our observation of the data, the changes within the frames at a distance equivalent to a stride of 4 from each other are insignificant. Each frame has a size of $3\times H\times W$, $H$, and $W$ stands for the height and width of the frame having three channels (RGB). We further divide these sub-sampled frames for a video into clips -- each clip containing 16 frames. We then randomly sample four clips to use during the pretraining phase. Before passing to the pretraining pipeline, the frames are resized to $224\times 224$.

\mypara{Token Generation}
We first divide a video $V$ of size $T\times 3\times H\times W$ into non-overlapping cubic tokens of size $2\times 3 \times 16 \times 16$. $T$ is the number of frames (temporal dimension), $H$ and $W$ are the height and width of the frames. Each frame has RGB channels. We use a 3D convolution of kernel size = $(2, 3, 16, 16)$,  stride $(2, 16, 16)$, and $d$ output channels. Using this 3d convolution layer, we generate a total of $N=\frac{T}{2}\times\frac{H}{16}\times\frac{W}{16}$ tokens, each of dimension $d$ ($d=384$ in our design) for every video. 
Next, we add the positional information to the tokens using the fixed 3D periodic positional encoding scheme introduced in \cite{vaswani2017attention}.

\mypara{Generating Object Localization Priors}
We utilize deep object detection networks as the region proposal network (RPN) to identify the potential GB region within a frame. The predicted bounding boxes are used as potential candidate regions containing the objects (malignancy). We used the public GBCU \cite{basu2022surpassing} dataset for training the object detectors. The GBCU dataset provides US images with regions-of-interest marked with bounding boxes. The training focuses on two classes: background and the GB region. We lower the confidence threshold of the predicted boxes to generate multiple candidate regions. These regions are used as priors in a masking token sampler to boost the masking probability of the tokens. If a token's spatial central point falls within the region prior, then its masking probability is inflated. To define a candidate region for an entire clip, we take the union of the candidate regions for each frame within the clip.

\mypara{Masked Token Sampling with Region Priors}
To generate the masking probabilities for the tokens, we follow \cite{adamae} and use an auxiliary network consisting of Multi-Head Attention (MHA) with a Linear and a Softmax ($\sigma$) layer following it. Given the embedded tokens $x \in \mathbb{R}^{N \times d}$, the probability scores $p \in \mathbb{R}^N$ over all tokens is generated as follows:
\begin{align}
z = \text{MHA}(x); \quad z \in \mathbb{R}^{N \times d} \\
p = \sigma(\text{Linear}(z)); \quad p \in \mathbb{R}^N
\end{align}
Region priors then boost the probability score as follows:
\begin{align}
\hat{p}_i = p_i + \pi_i 
\end{align}
If the $i$-th token spatially lies within the candidate regions, then we inflate the masking probability of the token by $\pi_i \in(0,\delta)$, where $\delta$ is a small fraction less than $0.25$. 
We then select without replacement a set of visible token indices $\mathcal{V} \in \{1,\ldots,N\}$ with the probability $(1-\hat{p}_i)$ for the $i$-th token. The set of masked token indices is given by $\mathcal{M} = \{1,\ldots,N\} \setminus \mathcal{V}$. The number of sampled visible tokens $N_v$ is computed based on a pre-defined masking ratio $\rho \in (0, 1)$ and equals $(1 - \rho)N$.

\mypara{Encoder}
For computational efficiency, only the visible (non-masked) tokens are passed to the encoder. The number of visible tokens is $N_v = (1-\rho)N$. We employed a vanilla ViT architecture with space-time attention \cite{timesformer}. The ViT encoder has a depth of 12 layers with 6 heads in each layer. The embedding dimension is 384. 

\mypara{Decoder}
The encoded visible tokens are appended with the masked tokens before passing to the decoder. The masked patches are learnable tokens that the decoder learns to reconstruct, guided by the MSE loss between the values of these tokens and their reconstructions. 
Usually, the decoder in an MAE is a shallow and narrow ViT. However, our experiments indicate that increasing the decoder depth can help in performance gain. We keep the decoder depth to 10 after grid searching for optimal depth. The decoder reconstructs the original video cube of size $\frac{T}{2}\times\frac{H}{16}\times\frac{W}{16}$ from the encoded and masked tokens.
%

\subsection{Training}
\mypara{Masking Reconstruction Loss} 
We have used the \emph{Mean Squared Error} loss (MSE) between the predicted and ground-truth RGB values of the masked tokens as the objective function to pretrain the MAE. The loss function is given as:
\begin{align}
   \mathcal{L}_{recon} = \frac{1}{|\mathcal{M}|}\sum_{i \in \mathcal{M}}||\hat{m}_i - m_i||_2 
\end{align}
Here $\hat{m}$ and $m$ denote the predicted token and the normalized ground-truth RGB values of the token. $|\mathcal{M}| = \rho N$ refers to the number of masked tokens. 

\mypara{Token Sampling Loss}
We use a token sampling loss, $\mathcal{L}_{sample}$, to train the sampling network that generates the sampling probability. We adapt the sampling loss proposed by AdaMAE \cite{adamae} and use maximization of the average reconstruction error to define the loss. The formulation of such a formulation is motivated by the expected reward maximization of the REINFORCE algorithm in RL. Here, the visible token sampling process is the \emph{action}, the MAE is the \emph{environment}, and the masked token reconstruction error is the \emph{return}. The reconstruction error is high in the high information regions as compared to the low information background regions. Thus, maximizing the expected reconstruction error would result in the network predicting a higher probability score for a high information region. The loss formulation is as follows:
\begin{align}
    \mathcal{L}_{sample} = - \sum_{i\in \mathcal{M}} \big(\log{\hat{p}_i} \cdot ||\hat{m}_i - m_i||_2 \big)
\end{align}
One key difference with the loss in AdaMAE is that the token probability in our formulation is augmented by the region priors, while AdaMAE uses a token probability for a distribution over the entire image. Thus, we obtain a more refined version of the adaptive token sampling. The log probability tackles the underflow and floating point errors. The gradient flow in the sampling network is kept independent from the ViT encoder and decoder of the main MAE. 

\begin{table*}[!t]
	\centering
    \small
	\begin{tabular}{llcccc}
		\toprule
		{\textbf{Group}} & {\textbf{Method}} & {\textbf{Backbone}} & {\textbf{Acc.}} &  {\textbf{Spec.}} & {\textbf{Sens.}} \\
		\midrule
        \multirow{2}{*}{Human Experts} 
		& Radiologist A  & -- & 0.786$\pm$0.134 & 1.000$\pm$0.000 & 0.672$\pm$0.201 \\
		& Radiologist B  & -- & 0.874$\pm$0.088 & 1.000$\pm$0.000 & 0.811$\pm$0.126 \\
		\midrule
		\multirow{10}{*}{Image-based} 
		& ResNet50 \cite{resnet} & CNN & 0.711$\pm$0.091 & 0.822$\pm$0.102 & 0.672$\pm$0.147 \\
		& InceptionV3 \cite{inception} & CNN & 0.734$\pm$0.089 & 0.953$\pm$0.072 & 0.647$\pm$0.107 \\
        & Faster-RCNN \cite{fasterrcnn} & CNN & 0.757$\pm$0.058 & 0.687$\pm$0.056 & 0.808$\pm$0.091 \\
		& EfficientDet \cite{efficientdet} & CNN & 0.789$\pm$0.084 & 0.761$\pm$0.099 & 0.828$\pm$0.061 \\
        \cmidrule{2-6}
        & ViT \cite{vit} & Transformer & 0.796$\pm$0.068 & 0.751$\pm$0.128 & 0.820$\pm$0.076 \\
		& DEIT \cite{touvron2021training} & Transformer & 0.829$\pm$0.034  & 0.787$\pm$0.154 & 0.845$\pm$0.058 \\
		& PVTv2 \cite{wang2021pvtv2} & Transformer & 0.831$\pm$0.041 & 0.857$\pm$0.167 & 0.834$\pm$0.068  \\
        \cmidrule{2-6}
        & GBCNet \cite{basu2022surpassing} & CNN & 0.840$\pm$0.105 & 0.843$\pm$0.204 & 0.843$\pm$0.072 \\
		& US-UCL \cite{basu2022unsupervised} & CNN & 0.808$\pm$0.127 & 0.871$\pm$0.217 & 0.776$\pm$0.109 \\
		& RadFormer (SOTA) \cite{basu2023radformer} & Transformer & 0.840$\pm$0.105 & 0.776$\pm$0.162 & 0.877$\pm$0.088 \\
		\midrule
		\multirow{5}{*}{Video-based} 
		%
        & Video-Swin \cite{videoswin} & Transformer & 0.925$\pm$0.053 & \textbf{1.000$\pm$0.000} & 0.903$\pm$0.085 \\
        & TimeSformer \cite{timesformer} & Transformer & 0.920$\pm$0.058 & 0.967$\pm$0.067 & 0.909$\pm$0.058 \\
        & VidTr \cite{vidtr} & Transformer & 0.924$\pm$0.038 & \textbf{1.000$\pm$0.000} & 0.800$\pm$0.072 \\
		& VideoMAEv2 \cite{videomaev2} & Transformer & 0.942$\pm$0.066 & 0.937$\pm$0.078 & 0.940$\pm$0.120 \\
		& AdaMAE \cite{adamae} & Transformer & 0.947$\pm$0.053 & 0.952$\pm$0.066 & 0.913$\pm$0.116 \\
		\cmidrule{2-6}
		& \myarch (Ours) & Transformer & \textbf{0.964$\pm$0.047} & 0.910$\pm$0.117 & \textbf{1.000$\pm$0.000} \\
		\bottomrule
	\end{tabular}
	\caption{The 5-fold cross-validation (Mean$\pm$SD) accuracy, specificity, and sensitivity of baselines and \myarch in detecting GBC from the US. \myarch achieves the best accuracy and perfect sensitivity, which is much desired for GBC detection. We also report how the expert radiologists perform in detecting GBC from the video dataset. The radiologists were blinded from accessing any patient-related data or clinical/ histopathological findings. The radiologists classified each video using the Gallbladder Reporting and Data Standard (GB-RADS) \cite{gb-rads-paper}. Our model outperforms human radiologists in detecting GBC from US videos. Recall that our ground truth labels are biopsy-proven. The performance of the expert radiologists in our study is comparable to literature \cite{gbc-lancet}. 
    }
	\label{tab:main}
\end{table*}

\begin{figure}
    \centering
    \includegraphics[width=\linewidth]{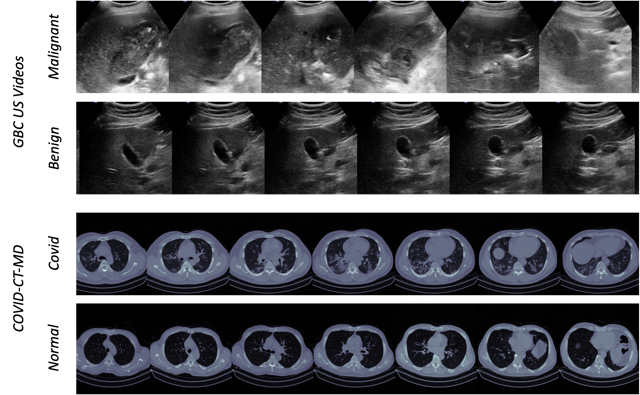}
    \caption{Sample video sequences from our US video dataset used for GBC detection, and the public COVID-CT-MD dataset \cite{covidctmd}. We show samples of both malignant and benign (non-malignant) sequences for GBC data. For the covid data, we show sample sequences for both Covid and non-Covid categories.}
    \label{fig:data_sample}
\end{figure}

\section{Dataset }
\subsection{Curated US Video Dataset for GBC Detection}
\mypara{Video Data Collection and Curation}
We utilized both the public Gallbladder US video dataset (GBUSV) \cite{basu2022unsupervised} and an additional set of US videos collected by our team of radiologists. The GBUSV dataset comprises 64 Gallbladder US videos, with 32 labeled as benign and another 32 labeled as malignant. To augment our dataset for the video-based \gbc detection task, we incorporated 27 additional US videos specifically depicting Gallbladder malignancy. 

We obtained video samples from patients referred to PGIMER, Chandigarh for abdominal US examinations targeting suspected Gallbladder pathologies. Each patient provided informed written consent during recruitment, and we ensure patient privacy by fully anonymizing the data \footnote{The institute Ethics Committee approved the study}. Patients were fasting for a minimum of 6 hours to ensure adequate distention of the \gb. Our team of radiologists employed a 1-5 MHz curved array transducer (C-1-5D, Logiq S8, GE Healthcare) for the scanning process. The scanning protocol covers the entire gallbladder (including fundus, body, and neck) and any associated lesions or pathologies. We cropped the video frames from the center to safeguard patient privacy and annotations. The processed frames have a size of 360x480 pixels. \cref{fig:data_sample} shows sample sequences from the dataset. 

\mypara{Annotation}
The video labels in GBUSV are already provided. For our additional videos, we relied on the biopsy reports for labeling. Additionally, two radiologists with 2 and 10 years of expertise in abdominal ultrasound (US), were consulted to draw bounding boxes covering the entire GB and the adjacent liver parenchyma in one frame in each video.  

\mypara{Dataset Statistics}
The dataset comprises 59 malignant and 32 non-malignant videos, collected from 41 malignant and 32 benign patients, respectively. In total, the dataset encompasses 21,955 frames, with 18,406 frames attributed to videos labeled as malignant. 

\mypara{Dataset Splits}
We report the 5-fold cross-validation metrics over the complete dataset for key experiments. The cross-validation splits were conducted on a patient-wise basis, ensuring that all videos of a particular patient appeared exclusively in either the training or the validation split during cross-validation.

\subsection{Public CT Dataset for Covid Detection}
We use the publicly available COVID-CT-MD dataset \cite{covidctmd} to assess the generality of our proposed method across different modalities and diseases. The COVID-CT-MD dataset contains lung CT scans of 169 (108 male and 61 female) confirmed positive COVID-19 cases, 76 (40 male and 36 female) normal cases and 60 (35 male and 25 female) Community-Acquired Pneumonia cases. All samples are annotated at the patient, lobe, and slice levels by three different radiologists. The authors used a Siemens SOMATOM Scope scanner to obtain the scans with the output size of the reconstructed images set to $512\times512$ pixels. Additionally, the dataset also contains clinical data, including the patient's age, gender, weight, symptoms, surgery history, follow-up and RT-PCR test reports. However, during our experiments, we did not use the clinical data. We used a stratified random 80:20 split to get the training and validation data.

\section{Implementation and Evaluation}

\mypara{Pretraining}
We implemented our experiments using PyTorch \cite{paszke2019pytorch}. We used Kinetics-400 pretrained weights for MAE weight initialization. Although there is a domain gap in natural and medical image data, studies show that pretraining on natural image data improves network performance on medical imaging tasks \cite{alzubaidi2020transferlearning, cheng2017transfer}.
We used the video sub-sampling scheme discussed in \cref{sec:method_subsample}. We apply random-resize cropping, random horizontal flipping, and random scaling as part of the data augmentations for pretraining. We chose ViT-S as the backbone. We use patch size of $2 \times 3 \times 16 \times 16$, resulting in $\frac{16}{2} \times \frac{3}{3}\times \frac{224}{16} \times \frac{224}{16} = 1568$ tokens for an input video of size $16\times3\times224\times224$.  The pretraining phase is trained with an AdamW optimizer with LR $0.0001$, layer decay $0.75$, and weight decay $0.05$, for minimizing the MSE loss over 300 epochs. The batch size was 2. Warm-up was done for 3 epochs with LR $0.001$.

\mypara{Fine-tuning}
\label{label:impl_ft}
For sub-sampling the videos during fine-tuning, a denser sample rate of 3 was used. We used 16 frames to constitute a clip. From each video, we sampled 5 clips uniformly. During inference, we predict the labels for each of the clips. If any of the clips is predicted as malignant, the entire video is labelled as malignant. We minimized a soft-target cross entropy loss using an AdamW optimizer with LR $1e-5$, layer decay $0.75$, and weight decay $0.05$ for 30 epochs. We used a batch size of 4. 

We have used a machine with an Intel Xeon Gold 5218@2.30GHz dual-core processor and 8 Nvidia Tesla V100 32GB GPUs for our experiments. 

\mypara{Evaluation Metrics}
We used video-level accuracy, specificity (true negative rate), and sensitivity (true positive rate/ recall) for assessing the video-based GBC identification. 

\begin{table}[!t]
    \small
	\centering
	\resizebox{ \linewidth}{!}{%
	\begin{tabular}{llccc}
		\toprule
		{\textbf{Group}} & {\textbf{Method}} & {\textbf{Acc.}} &  {\textbf{Spec.}} & {\textbf{Sens.}} \\
		\midrule
		\multirow{5}{*}{Image-based} 
		& ResNet50 \cite{resnet} & 0.721 & 0.739 & 0.711 \\
		& InceptionV3 \cite{inception} & 0.672 & 0.739 & 0.632 \\
        %
		%
        \cmidrule{2-5}
        & ViT \cite{vit} & 0.770 & 0.783 & 0.763 \\
		& DEIT \cite{touvron2021training} & 0.770 & 0.696 & 0.816 \\
		\midrule
		\multirow{3}{*}{Video-based} 
		%
		%
        & TimeSformer \cite{timesformer} & 0.700 & 0.739 & 0.474 \\
        %
		%
		& VideoMAE \cite{videomaev2} & 0.852 & \textbf{0.956} & 0.789 \\
		%
		%
		\cmidrule{2-5}
		& \myarch (Ours) & \textbf{0.885} & 0.895 & \textbf{0.869} \\
                
		\bottomrule
	\end{tabular}
	}
	\caption{The performance comparison in terms of accuracy, specificity, and sensitivity of baselines and \myarch for detecting COVID from CT \cite{covidctmd}. CT-slices are analogous to the video frames, and thus, video-based detection methods are applicable to CT modality as well. Our proposed method consistently outperforms the SOTA baselines on the COVID detection task, establishing the generality and applicability of our method across two different medical imaging modalities -- US and CT. }
	\label{tab:covid}
\end{table}

\begin{figure}[t]
    \centering
    \includegraphics[width=\linewidth]{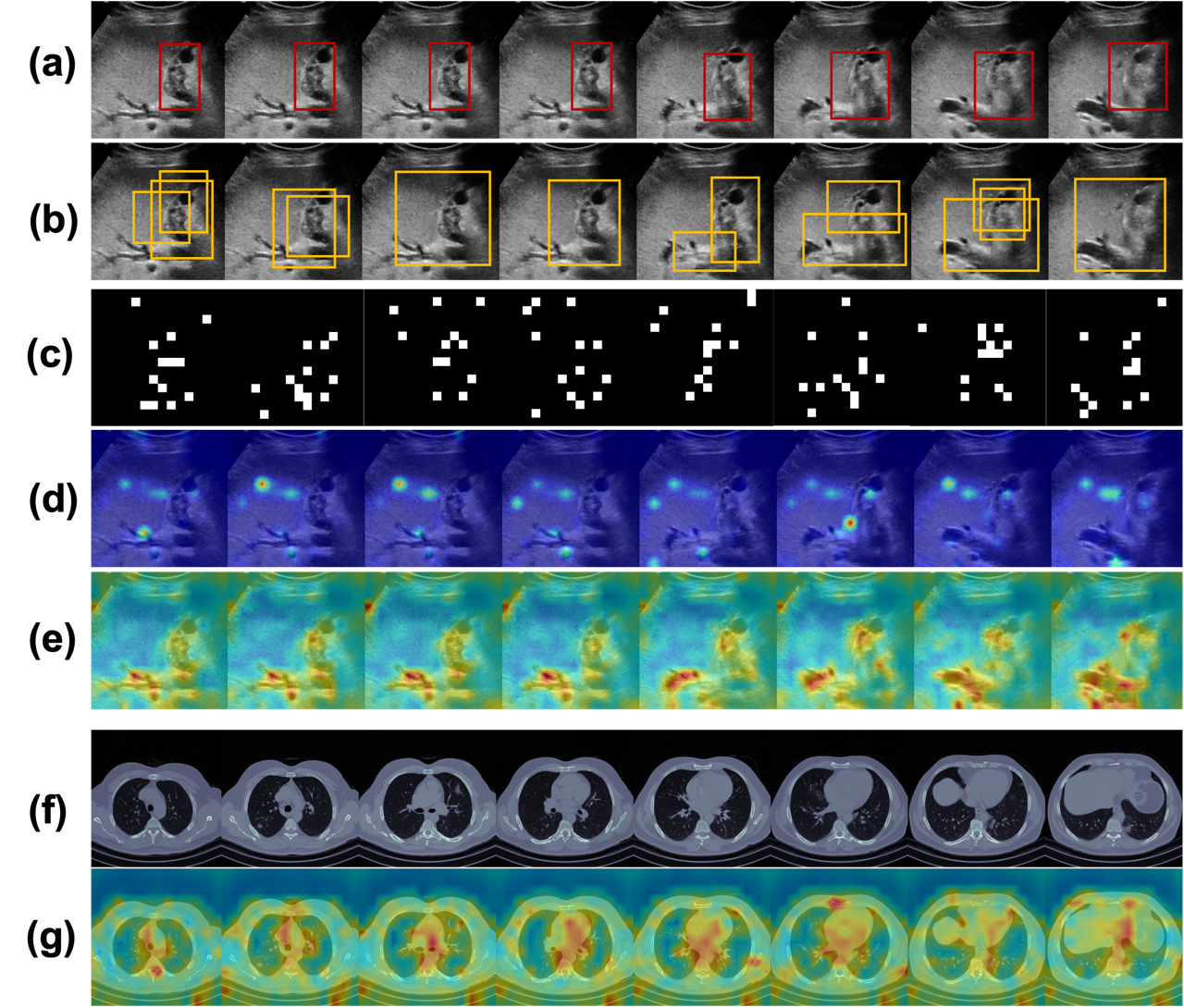}
    \caption{Visual demonstration of the benefit of using the \myarch method. (a) Original frames from a US video sequence exhibiting GB malignancy. ROI is drawn in red. (b) Candidate regions as prior (in yellow). (c) Masking by \myarch. (d), (e) Attention visualization for the downstream malignancy detection for VideoMAE and \myarch, respectively. For \myarch, the attention is well guided to the key regions containing the malignancy, as opposed to VideoMAE. (f) CT Slices of a sample Covid patient. (g) Attention visualization of \myarch.}
    \label{fig:qualitative}
\end{figure}

\section{Experiments and Results}
\subsection{Efficacy of \myarch over SOTA Baselines}
We explore the GBC classification performance on US videos for five SOTA video classification methods, namely Video-Swin \cite{videoswin}, TimeSformer \cite{timesformer}, VidTr \cite{vidtr}, VideoMAEv2 \cite{videomaev2}, and AdaMAE \cite{adamae}. 

In addition, we have also explored three SOTA techniques \cite{basu2022surpassing, basu2022unsupervised, basu2023radformer} that are specialized for GBC detection on US images. Apart from these specialized models, we analyze the performances of popular image-centric CNN-based classifiers \cite{resnet,inception} and detectors \cite{fasterrcnn, efficientdet}. We also look into three popular Transformer-based classifiers -- ViT \cite{vit}, DEiT \cite{touvron2021training}, and PvT \cite{wang2021pvtv2} for GBC detection.

\mypara{Using Image-based Methods for Video Classification}
We use the same video sub-sampling scheme used during the fine-tuning phase (ref. \cref{label:impl_ft}) of the \myarch to get the frames and clips. We then use the image-centric methods to predict the labels for each frame in the clips. If the majority of the frames in a clip are predicted as malignant, then the clip is predicted as malignant. If any clip within a video is predicted as malignant, the overall video is categorized as malignant. The image-based methods were pretrained on the public GBCU \cite{basu2022surpassing} dataset. 

\mypara{Quantitative Analysis}
We show the 5-fold cross-validation performance in terms of accuracy, specificity, and sensitivity for the baselines and the proposed \myarch in \cref{tab:main}. Clearly, the video-based techniques trump the image-centric SOTA methods of GBC detection, supporting our recommendation of a paradigm shift to video-based classification for the problem. Additionally, we see the effectiveness of the \myarch in detecting GBC.

\mypara{Qualitative Analysis}
We show the qualitative analysis in \cref{fig:qualitative}. The random masking by VideoMAE does not adequately mask the high-information malignant region. In contrast, the region prior guided \myarch generates stronger masking for learning the malignant representation by biasing the masking towards the malignancy localization region. We visualize the attention rollout during the downstream task. Clearly, \myarch's attention regions highlight semantically more meaningful areas, such as the gallbladder boundary and anatomical structures, compared to VideoMAE.

\subsection{Generality of the Proposed Method}
We explored the generality of the proposed \myarch method on the task of Covid detection from a publicly available CT dataset \cite{covidctmd}. \cref{tab:covid} shows that \myarch achieves much better accuracy, specificity, and sensitivity, indicating the superiority of the disease representation learning capability of \myarch. The applicability of \myarch on two distinct tasks - 1) GBC detection from US videos, and 2) Covid detection from CT - establishes the generality of the method across two diagnostic modalities, and diseases.

\begin{figure}[t]
	\centering
	\begin{subfigure}[b]{0.49\linewidth}
        \centering
        \includegraphics[width=\linewidth, height=10em]{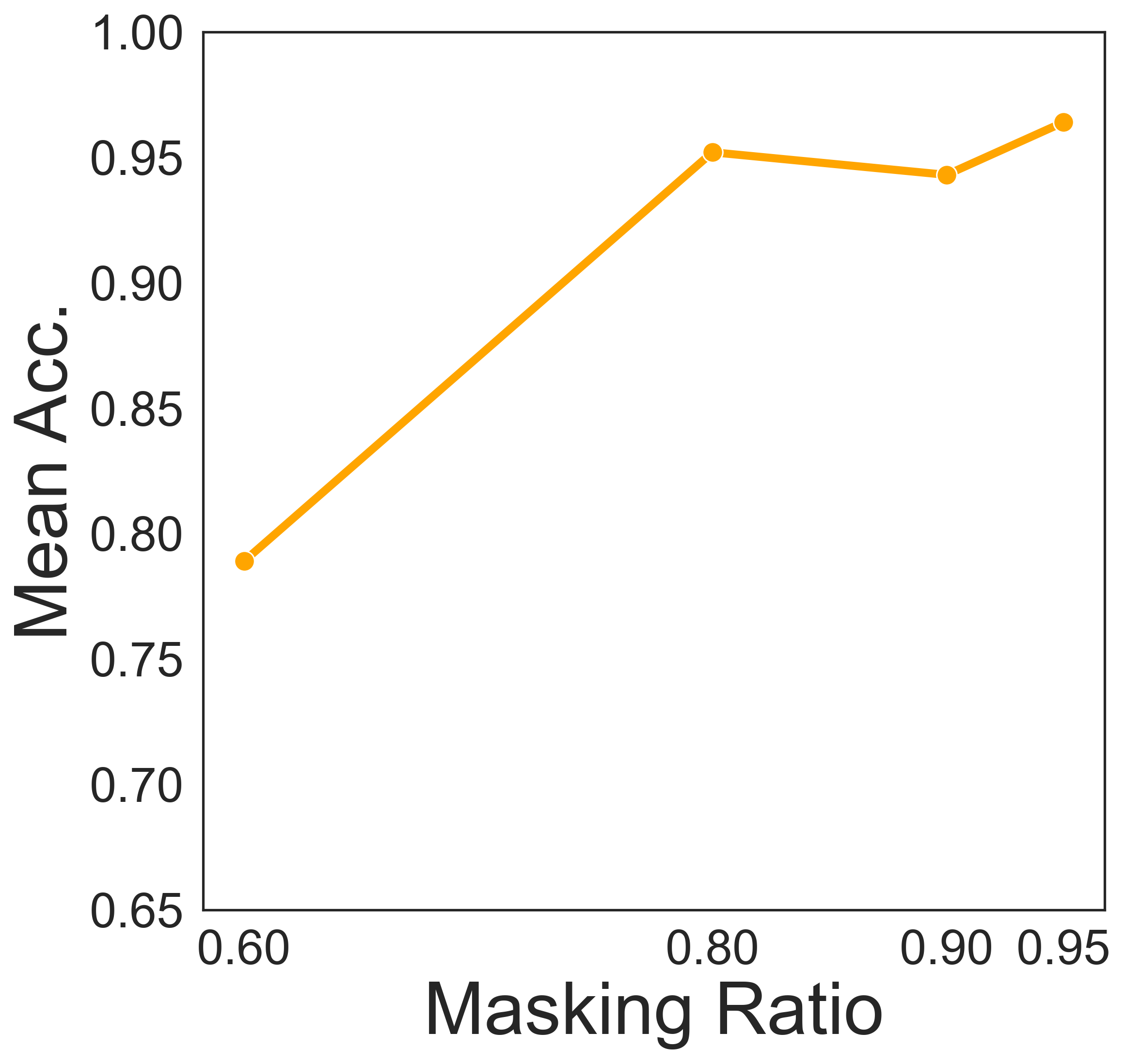}
        \caption{}
        \label{fig:ablation_mr}
    \end{subfigure}
	\begin{subfigure}[b]{0.49\linewidth}
		\centering
		\includegraphics[width=\linewidth, height=10em]{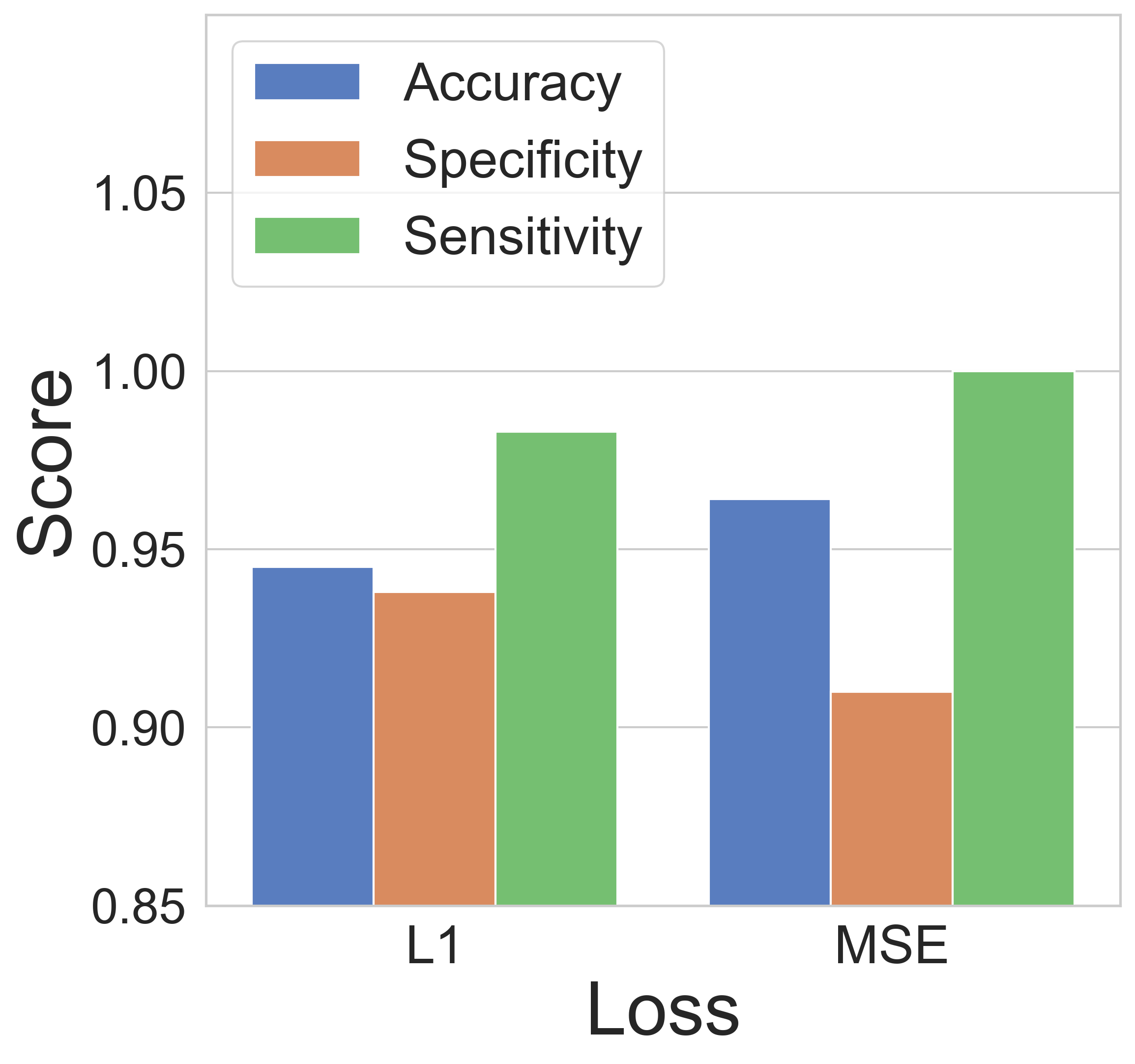}
		\caption{}
		\label{fig:ablation_loss}
	\end{subfigure}
 
	\begin{subfigure}[b]{0.49\linewidth}
		\centering
		\includegraphics[width=\linewidth, height=10em]{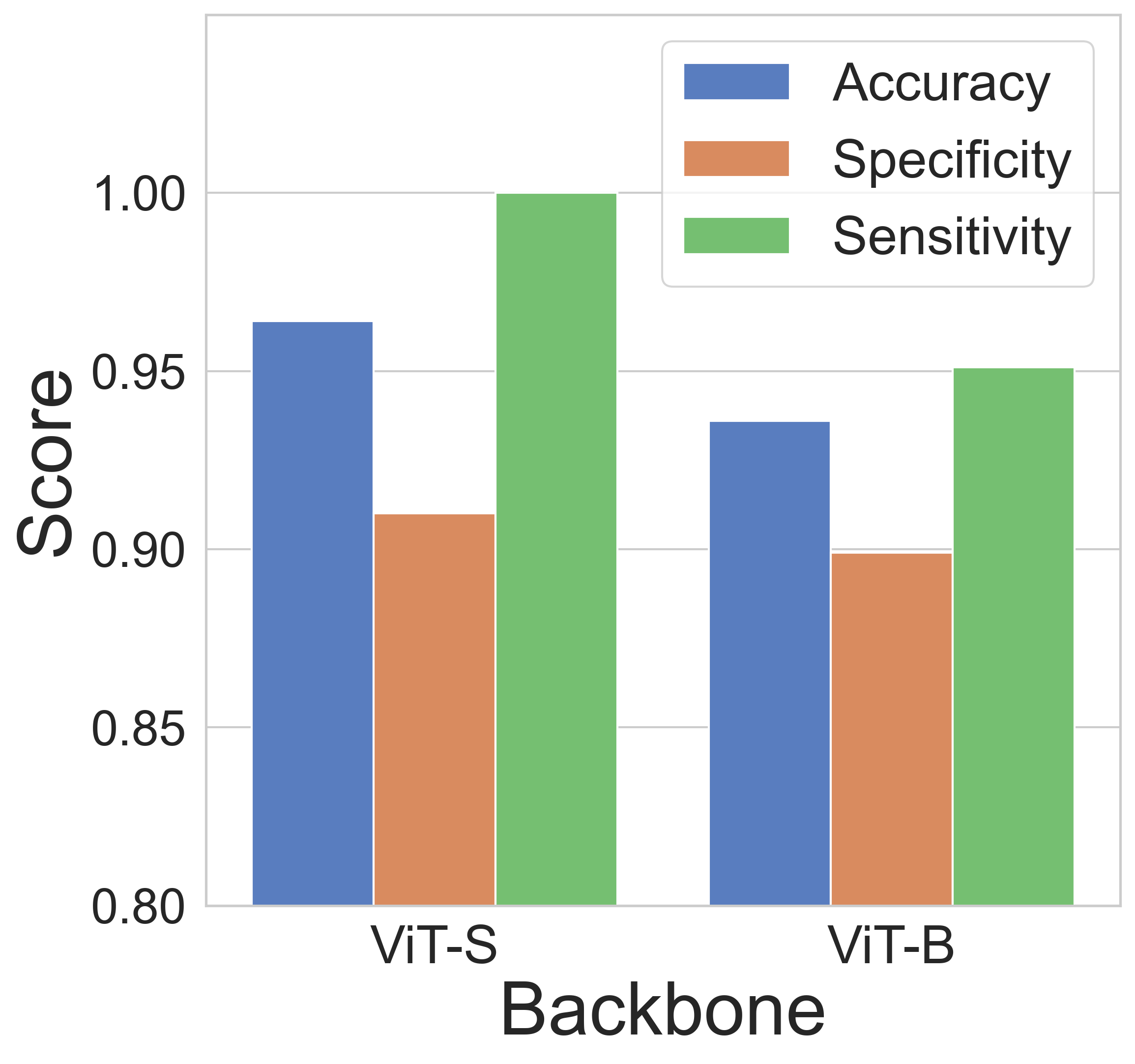}
		\caption{}
		\label{fig:ablation_enc}
    \end{subfigure}	
    \begin{subfigure}[b]{0.49\linewidth}
		\centering
		\includegraphics[width=\linewidth, height=10em]{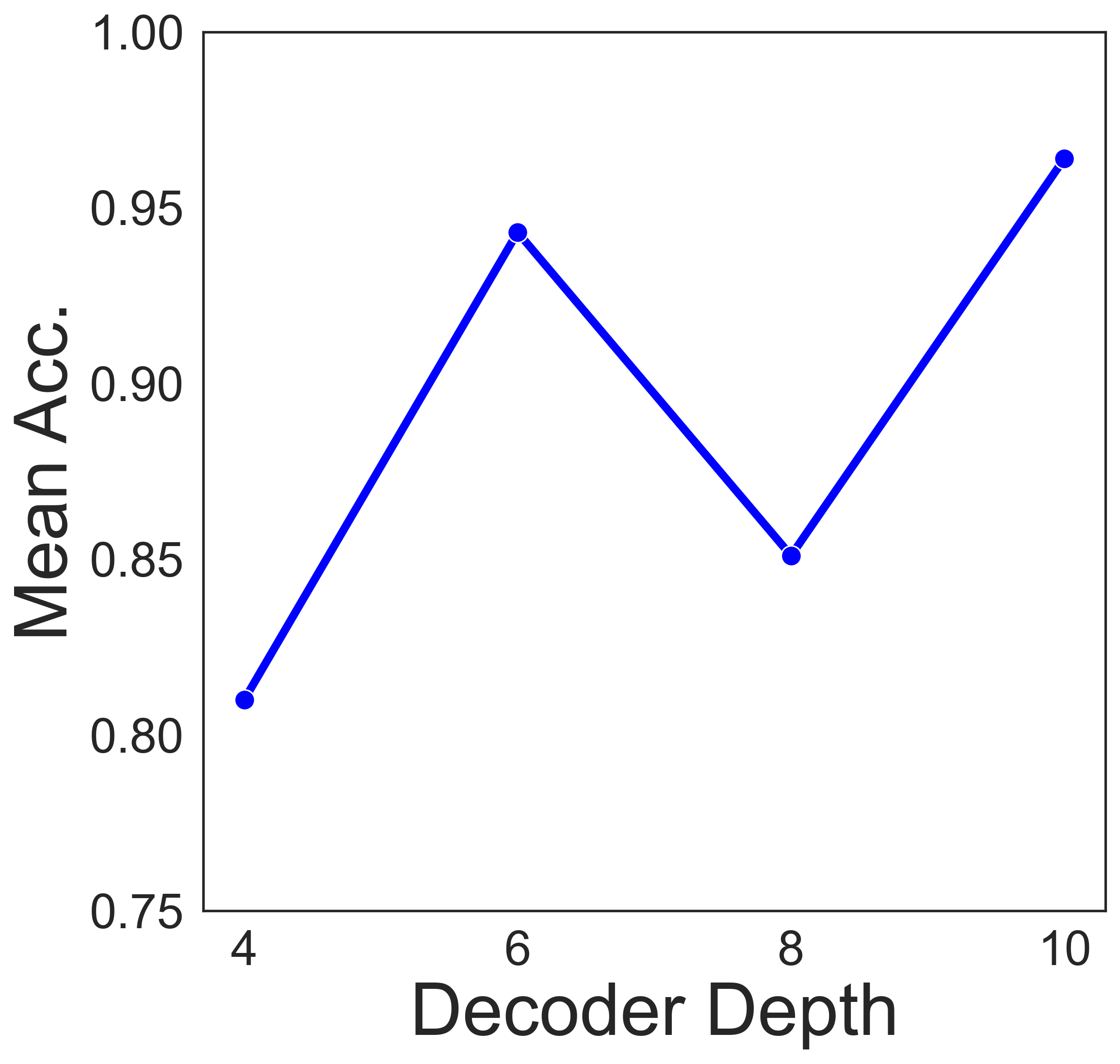}
		\caption{}
		\label{fig:ablation_dec}
	\end{subfigure}
	\caption{Ablation study. We report the mean scores over 5-fold cross-validation for GBC detection. (a) Effect of varying the masking ratio ($\rho$) on accuracy. (b) Effect of varying the reconstruction loss - L1 vs. MSE - for SSL pretraining. Training with MSE yields 2.1\% better accuracy. (c) Performance for different backbones. (d) Effect of varying the decoder depth. }
	\label{fig:ablation}
\end{figure}

\begin{figure}
    \centering
    \includegraphics[width=\linewidth]{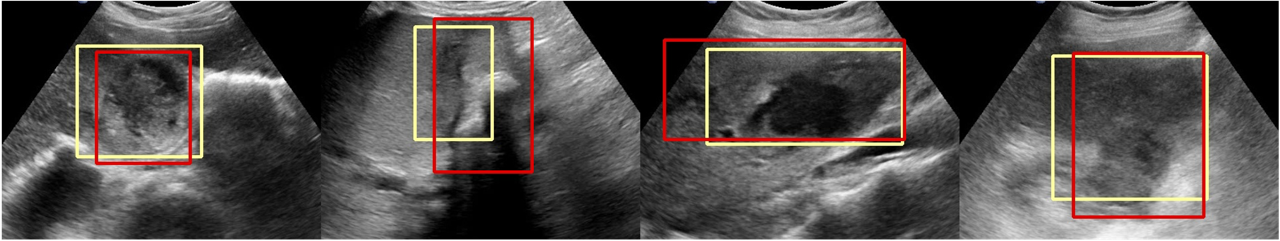}
    \caption{Visuals of candidate regions. Red -- malignant regions identified by radiologists. Yellow -- candidate object localization generated by the RPN.}
    \label{fig:roi}
\end{figure}

\subsection{Ablation Study}
We performed the ablation study on the FocusMAE with the ViT backbone on the US Video dataset. 

\mypara{Masking Ratio}
\cref{fig:ablation_mr} shows how the masking ratio $\rho$ influences the performance of \myarch. For \myarch, 95\% masking ratio achieves the best accuracy of 96.4\%. VideoMAE uses a 90\% masking ratio with random tube-based masking. The region-prior guided approach helps \myarch to sample more informative tokens with lower redundancy than VideoMAE.

\mypara{Reconstruction Loss}
We examined the effect of varying the reconstruction loss function in our study. We experimented with two variants: Mean Absolute Error or L1-loss, and Mean Squared Error (MSE). The results, shown in \cref{fig:ablation_loss}, indicate that using MSE loss during pretraining produces slightly better performance in terms of accuracy. Models trained with MSE loss demonstrated 2.1\% higher mean accuracy compared to those trained with L1 loss.

\mypara{Encoder Backbone}
\cref{fig:ablation_enc} demonstrates the effect of ViT variants on the token encoding task. We experimented with ViT-S and ViT-B. We observe that larger backbones do not perform well for our data, indicating potential over-fitting.

\mypara{Decoder Depth}
We experiment with the number of decoder blocks and present the result in \cref{fig:ablation_dec}. We see performance gain when the decoder depth is varied from 4 to 6. However, there is a drop in performance when the decoder depth is increased to 8. The observation is consistent with \cite{videomae, adamae}. Interestingly, when we increased the depth further, we saw an increase in accuracy, which indicates that the decoder can benefit from increasing the depth and need not necessarily be a shallow network. 

\subsection{Analysis on Candidate Region Selection}
\cref{fig:roi} shows sample object region localization of the RPN. We adopted a FasterRCNN-based RPN for generating the candidate regions for using as priors in \myarch. The RPN achieves mIoU of 0.712 with a recall rate of 0.994.

\section{Conclusion}
This study addresses the limitations of current US image-based GBC detection techniques, emphasizing the need for a paradigm shift towards US video-based approaches. Our novel design, named \myarch, strategically biases masking token selection from high-information regions and learns quality representations of GB malignancy. \myarch achieves state-of-the-art results on US video-based GBC detection.  
We hope that our work will 
spark interest in the challenging problem of GBC detection from US videos. Moreover, we showcase the generality of \myarch by applying it successfully to a public lung CT-based Covid detection task, demonstrating its applicability across two modalities and diseases. This suggests that \myarch could find broader use-cases in future, marking it a promising step towards versatile diagnostic solutions.

\vspace{0.4em}
\mypara{Acknowledgments}
{\small
Authors thank Dr. Shravya Singh and Dr. Ruby Siddiqui for data annotation, and Dr. Pratyaksha Rana for reading the cases.  
This work was partially supported by the CSE Research Acceleration Fund of IIT Delhi. 
}

\clearpage

{\small
\bibliographystyle{ieee_fullname}
\bibliography{main}
}

\clearpage
\appendix
\beginsupplement

\section*{Supplementary Material}

\maketitle

\appendix
\beginsupplement

\section{Region Selection Network}
We have experimented with multiple deep detectors to localize the candidate regions. We haven't used the bounding box annotations in the video for training the candidate region networks. Instead, we used the public GBCU dataset to pretrain the detectors for localizing the malignancy. We then lowered the threshold to generate multiple candidate regions for the video frames used in the \myarch experiments. 

To calculate precision and recall in the GB localization phase, following the recommendation of \cite{ribli2018detecting}, we determine a predicted region as a true positive if its center falls within the bounding box of the ground truth region. Conversely, if the center is outside the bounding box, we categorize the prediction as a false positive attributed to localization error. \cref{tab:rpn} shows the mIoU and the recall for the different candidate region detectors.

\begin{figure}[ht]
    \centering
    \includegraphics[width=\linewidth]{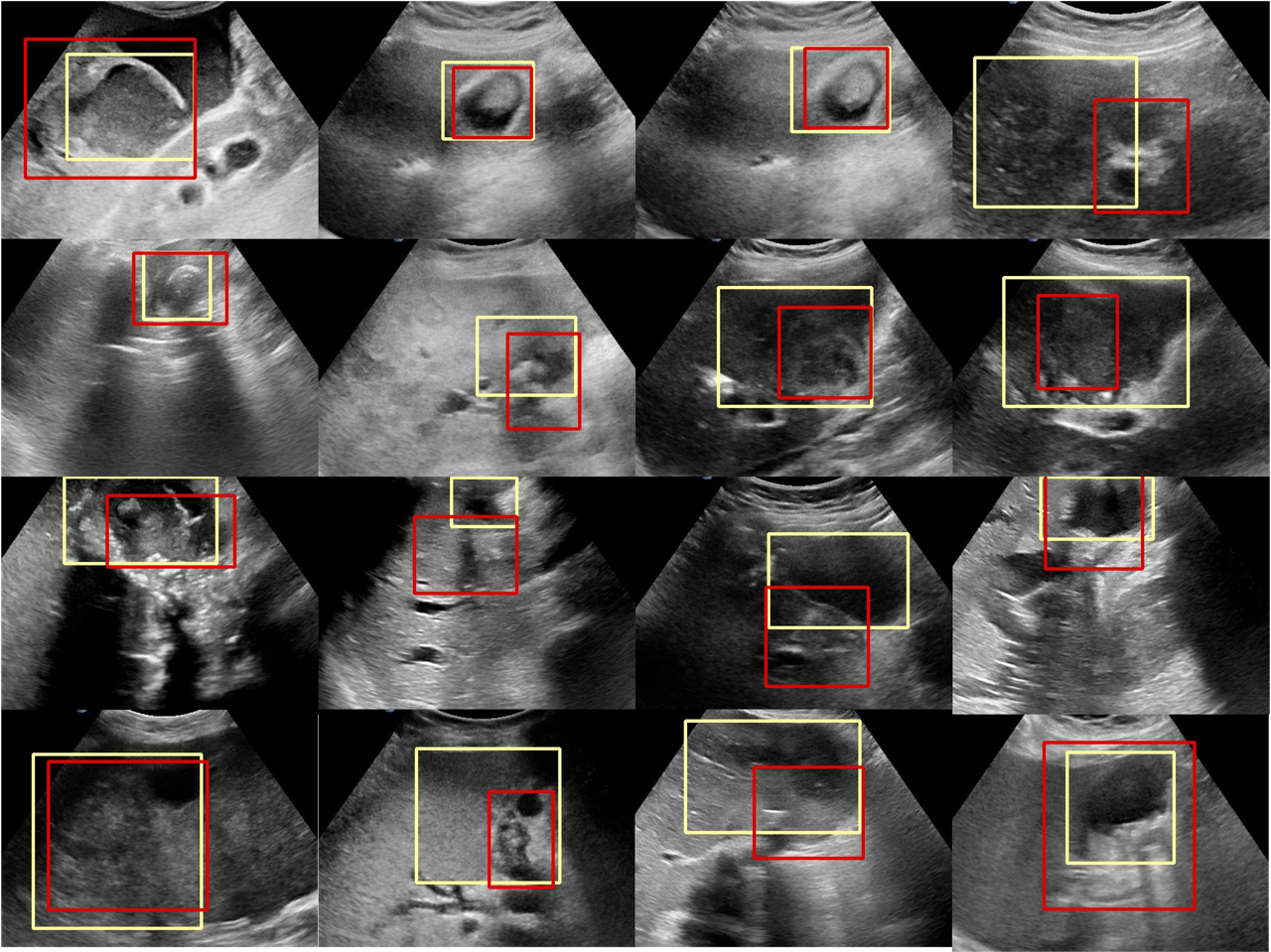}
    \caption{Sample candidate bounding boxes generated by the region selection network.}
    \label{fig:rpn_vis}
\end{figure}

\cref{fig:rpn_vis} shows sample object region localization of the RPN. We adopted a FasterRCNN-based RPN for generating the candidate regions for using as priors in \myarch as the detector achieves the best recall rate.

\begin{table}[ht]
	\centering
	\small
    \begin{tabular}{@{}lccc@{}}
    \toprule[1pt]
    \textbf{Model} & \textbf{mIoU} & \textbf{Precision} & \textbf{Recall} \\
    \midrule[0.5pt]
    Faster-RCNN & 0.712 &  0.952 & 0.994 \\
    YOLO  & 0.767 & 0.979 & 0.962 \\
    CentripetalNet & 0.614 & 0.947 & 0.909 \\
    Reppoints & 0.682 & 0.942 & 0.997 \\
    DETR & 0.724 & 0.962 & 0.988 \\
    \bottomrule[1pt]
    \end{tabular}
	\caption{Comparison of the candidate region selection models.}
\label{tab:rpn}
\end{table}

\section{Region Selection Network Implementation}
We adopted the Faster-RCNN \cite{fasterrcnn} model for candidate region selection. A frozen Resnet50 Feature Pyramid backbone is used. The input size was $800 \times 1333 \times 3$. We used a SGD optimizer with LR = 0.005, momentum = 0.9, and weight decay = 0.0005. We used a batch size of 16 and trained for 60 epochs on the GBCU dataset.

\section{Visualization}
\begin{figure*}[ht]
    \centering
    \includegraphics[width=\textwidth]{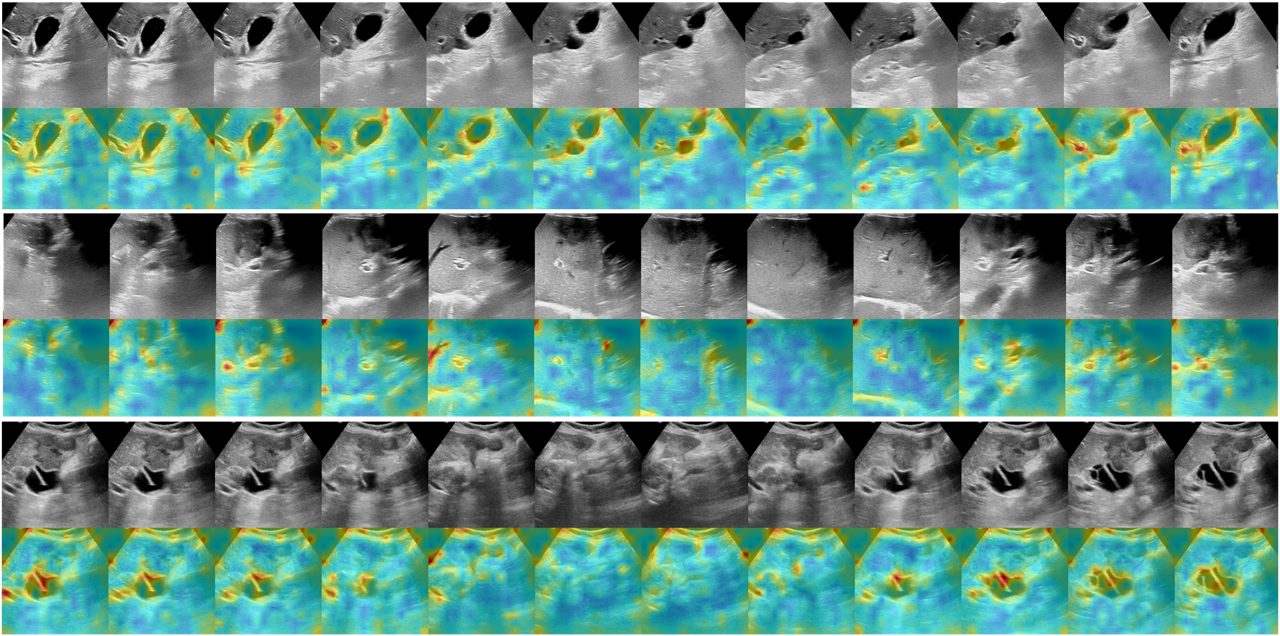}
    \caption{Attention visuals for \myarch for the GBC detection task on the US videos. We show three different maligant video samples. For each video sample, the upper row shows the sequence with the original frames, and the lower row shows the attention on the frames.}
    \label{fig:visual_supp_gbc}
\end{figure*}

\begin{figure*}[t]
    \centering
    \includegraphics[width=\textwidth]{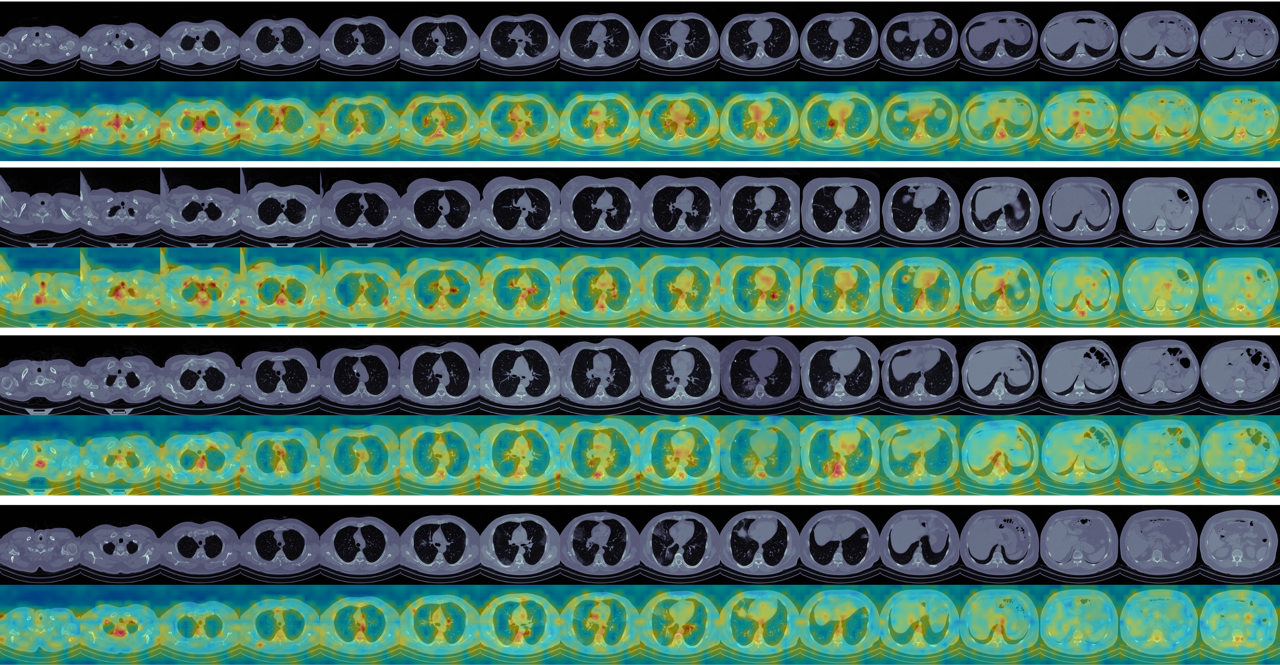}
    \caption{Attention visuals for \myarch for COVID detection from CT images. We show four COVID CT samples. For each  sample, the upper row shows the sequence with the original CT slices, and the lower row shows the attention on these slices.}
    \label{fig:visual_supp_covid}
\end{figure*}

\cref{fig:visual_supp_gbc} and \cref{fig:visual_supp_covid} show the attention visuals for the proposed \myarch method on additional data samples. Evidently, \myarch is able to attend the salient regions for disease detection.

\begin{table*}[t]
\centering
\resizebox{ \linewidth}{!}{%
\begin{tabular}{p{0.12\linewidth}p{0.38\linewidth}p{0.08\linewidth}p{0.2\linewidth}p{0.05\linewidth}p{0.06\linewidth}}
\toprule[1.5pt]
\textbf{Model} & \textbf{Description} & \textbf{Input Size} & \textbf{Optimizer} & \textbf{Batch size} & \textbf{Epochs/ Steps}
\\ \midrule[0.75pt]

VideoMAEv2 \cite{videomaev2} & Vision transformer (ViT) backbone, with random masked auto-encoders. Masking in both encoder and decoders. All attention-based layers were trainable. ViT base model used for inference. & $3\times16\times224\times224$ & AdamW, LR = 7e-5, momentum = 0.999 ,weight decay = 0.1  & 4 & 30 epochs\\ \midrule



%
TimeSformer \cite{timesformer} & Vision transformer based space time atention. Divided space-time attention configuration used. ViT base model used for inference.  & $3\times8\times224\times224$ & SGD, LR = 0.005, weight decay=1e-4, momentum=0.9 & 8 & 25 epochs\\ \midrule
VideoSwin \cite{videoswin} & Pretrained on ImageNet-1K. SwinTransformer3D based backbone. All layers were trainable & $3\times8\times224\times224$ & SGD, LR = 0.01, weight decay=1e-4, momentum=0.9 & 4 & 30 epochs \\ \midrule
AdaMAE \cite{adamae} & Vision transformer (ViT) backbone, with adaptively masked auto-encoders. The model embedding size provided by the authors is 768; we have pre-trained the 384 version to allow for a better fit to our data. Masking in only encoders. All attention-based layers were trainable. ViT base model used for inference  & $3\times16\times224\times224$ & AdamW, LR = 1e-6, weight decay=0.9, momentum=0.99 & 2 & 10 epochs \\ \midrule
VidTr \cite{vidtr} & Transformer-based video classification with separable attention. ViT-B backbone. All layers were trainable. & $3\times16\times224\times224$ & SGD, LR = 3e-4, weight decay=1e-5, momentum=0.9 & 2 & 40 epochs \\ 
%
\bottomrule
\end{tabular}
}
\caption{Implementation details for the different video-based baseline networks used for US video-based classification of Gallbladder Cancer. All details are for finetuning on the GB US video dataset. Pretraining details for VideoMAE and AdaMAE are already discussed.}
\label{tab:configs}
\end{table*}

\section{Baseline Implementation Details}
\label{supp:impl}
\cref{tab:configs} lists the configurations of all baseline models used in this study. We trained our models on 4 Nvidia Tesla V100 32GB GPUs. The table includes a brief description of the model, input sizes, optimizer parameters, other relevant hyper-parameters such as learning rate, weight decay, momentum, batch size, and the number of training epochs for the network.  

For VideoMAEv2 pretraining, we used a Vision transformer (ViT) backbone, with random masked auto-encoders. Masking was done in both encoder and decoders. 
All attention-based layers were trainable. We used the ViT-S model. The input size was $3\times16\times224\times224$. We initiated the ViT weights with the Kinetics-pretrained VideoMAE weights. We have optimized the MSE loss for original and reconstructed masked patches on the GBC US Video dataset using an AdamW optimizer with LR = 1e-4 and momentum = 0.95. We used a batch size of 32 and trained for 1200 epochs.

AdaMAE pretraining was an adoption of the VideoMAE pretraining procedure. We used the ViT-S backbone, with adaptively masked auto-encoders. We have pre-trained the model with embedding dimension 384 to allow for a better fit to our data. We have used masking in only encoders. All attention-based layers were trainable. Similar to VideoMAE, we initialized the weights with the Kinetics preained AdaMAE weights. We used the MSE loss and used an AdamW optimizer with LR = 1e-4 and momentum = 0.95. The input sizes are  $3\times16\times224\times224$. We used batch size of 8 and pretrained for 500 epochs.

\section{Clip-level Statistics}
We have a total of 484 clips sub-sampled from the 91 videos at the fine-tuning stage. Out of these, 320 clips were from the malignant videos, and contain the malignant label as per the positive biopsy reports. All clips of a malignant video is given the malignant label. Radiologists identified 199 clips out of these 320 to be malignant. At a frame-level, radiologists identified 3212 frames exhibiting signs of malignancy.

		



		

\end{document}